\documentclass[twocolumn,amsmath,amssymb,aps,prl]{revtex4-2}
\usepackage{CJK}
\usepackage{xcolor}
\usepackage{graphicx}
\usepackage{subfigure}
\usepackage{dcolumn}
\usepackage{bm}
\usepackage{tabularx}
\usepackage{amsmath}
\usepackage[colorlinks,linkcolor=blue, urlcolor=blue, anchorcolor=blue, citecolor=blue]{hyperref}
\usepackage[title]{appendix}

\usepackage{ulem}

\begin{document}
\title{Superradiance fingerprints of moir\'e exciton-biexciton condensates: path-entangled biphoton vs. supercurrent-enabled one-photon channel}
\author{Haochen Wang$^{1,2,3}$}
\author{Shengxian Li$^{4,5,6}$}
\author{Hongyu Lu$^{1,2,3}$}
\author{Mingxing Chen$^{4,5,6}$}
\author{Wang Yao$^{1,2,3}$}
\email{wangyao@hku.hk}
\address{$^1$ New Cornerstone Science Laboratory, Department of Physics, The University of Hong Kong, Hong Kong, China}
\address{$^2$ HK Institute of Quantum Science and Technology, The University of Hong Kong, Hong Kong, China}
\affiliation{$^3$ State Key Laboratory of Optical Quantum Materials, The University of Hong Kong, Hong Kong, China}
\address{$^4$ Key Laboratory for Matter Microstructure and Function of Hunan Province, Key Laboratory of Low-Dimensional Quantum Structures and Quantum Control of Ministry of Education, Changsha 410081, China}
\address{$^5$ Hunan Research Center of the Basic Discipline for Quantum Effects and Quantum Technologies, Changsha 410081, China}
\address{$^6$ School of Physics and Electronics, Hunan Normal University, Changsha 410081, China}

\date{\today}
\begin{abstract}
{\bf 
Hybrid moir\'e excitons form a dense array of electrically tunable single-photon emitters, entailing cooperative optical responses with intriguing consequences. With the strong dipole-dipole interaction, they also realize Bose-Hubbard quantum simulators of emergent correlated matter such as Bose-Einstein condensates (BEC). Here we reveal the characteristic cooperative radiation from condensates of hybrid moir\'e excitons in twisted MoTe$_2$ bilayers. In each individual moir\'e trap, the layer hybridization selection rule dictates that the ground-state exciton is a quantum emitter of orbital-angular-momentum photons,
with an exceptionally long radiative lifetime.
Ground-state emission becomes pronounced only upon forming condensates, where inter-sublattice dipole attraction and intra-sublattice repulsion together lead to the coexistence of exciton and biexciton BEC with comparable weights. 
The biexciton (molecular-type) BEC decays through superradiance of path-entangled biphotons, whereas the exciton (atomic-type) BEC is dark at rest and becomes superradiant when it carries a supercurrent.
Our findings not only uncover a novel source of entangled light but also establish unique fingerprints for exploring the cold chemistry of atomic-molecular condensates with moir\'e excitons.}
\end{abstract}
\maketitle


Moir\'e superlattices in van der Waals structures of 2D semiconductors realize nanodot confinements of excitons with a  periodicity up to two orders of magnitude smaller than the optical wavelength~\cite{Hongyi2017scia,Seyler2019MoirExcitons,Jin2019MoirExcitons,Tran2019MoirExcitons,Alexeev2019HybridizedExcitons,Baek2020SA,Zhao2023NN}. These moir\'e excitons are generally hybrids of intralayer and interlayer components, with the degree of hybridization set by the local band alignment between the layers~\cite{Alexeev2019HybridizedExcitons,
RuizTijerina2019PRB,Shimazaki2020,Hongyi2021PRX,Zhao2024NanoL}. Light coupling is primarily determined by the intralayer component. The uniformity and deep-subwavelength periodicity of the exciton emitter array underlie the cooperative nature of the light-matter interaction~\cite{CamachoGuardian2022Nonlinearities,Huang2025CollectiveOptics,Kumlin2025Superradiance,Devenica2026CollectiveEmission,Xu2026CooperativeOptics}. The permanent dipole associated with the interlayer component underlies electric tunability of the emitter array~\cite{Xu2026CooperativeOptics}, as well as strong dipole-dipole interactions for realizing Bose-Hubbard quantum simulators of correlated bosonic matter~\cite{Hongyi2021PRX,Li2020DipolarInteractions,Gotting2022PRB,Park2023NP,Xiong2023Science,Gao2024NC,Zhai2025TwistronicsReview}.
Notably, cooperative light emissions can provide information on excitons' subwavelength spatial configurations, including the lattice geometry~\cite{Huang2025CollectiveOptics}, and density order at partial filling~\cite{Kumlin2025Superradiance,Devenica2026CollectiveEmission}. 
These moir\'e exciton platforms significantly expand the parameter regime for exploring cooperative quantum optics from bosonic quantum emitters previously pursued primarily with atomic ensembles~\cite{GrossHaroche1982Superradiance}. 
Existing studies, on the other hand, are limited to the realm where each quantum emitter is a simple dipole source of unstructured light.

Twisted bilayer MoTe$_2$ not only provides a magic platform for the groundbreaking discovery of fractional Chern insulators~\cite{Cai2023FQAH,Zeng2023FCI,Park2023FQAH,Xu2023FQAH}, and for unconventional superconductivity~\cite{Xu2025Superconductivity}, but also hosts unusual moir\'e excitons for cooperative quantum optics and correlated boson matter~\cite{Hongyi2021PRX}.
Interlayer excitons are strongly confined to the MX and XM stacking regions by oppositely oriented moir\'e electrical polarization~\cite{HongyiNSR2020,zhang_polarization-driven_2024,Falko2020prl}.
The C$_3$ rotational symmetry dictates that the moire exciton ground state is formed by the S-wavepacket interlayer component hybridizing a P-wavepacket intralayer component~\cite{Hongyi2021PRX}. This unusual configuration leads to the emission of structured photons, but the radiative lifetime can be extremely long in the absence of collective enhancement. Consequently, these long-lived excitons provide a bosonic quantum many-body simulator featuring an electric dipole-dipole interaction that is strongly attractive between nearest-neighbor MX and XM sites and repulsive within each sublattice. Moir\'e excitons can therefore exist either as monomers or as bound dimers (quadrupolar biexcitons) on the MX–XM bonds~\cite{NSR_Wang_2026}.

Here we uncover the cooperative radiation properties of hybrid moir\'e excitons and their condensates in twisted MoTe$_2$. The lowest energy exciton in an individual moir\'e trap has a non-degenerate, valley-coherent configuration, which emits an orbital-angular-momentum (OAM) photon.
The ground state emission becomes pronounced only upon forming condensates, where the interplay of inter-sublattice attraction and intra-sublattice repulsion yields coupled exciton and biexciton BEC of comparable weights. Cooperative electrodynamics of the structured quantum emitters leads to unusual radiative behavior: radiation from the zero-momentum exciton condensate is completely quenched, while the biexciton condensate decays via superradiant emission of momentum-entangled photon pairs. An in-plane gradient of the displacement field couples to the dipolar exciton condensate (but not the quadrupolar biexciton condensate), accelerating it to finite momentum, at which the exciton condensate becomes superradiant, emitting highly directional photons with intensity set by the supercurrent density. These distinct superradiant and electrical responses not only enable tunable quantum light sources but also provide a pathway to explore the cold chemistry of exciton-biexciton condensates via their unique optical fingerprints.

\textit{Hybrid moire excitons in structurally relaxed twisted bilayer}. Atomic reconstruction produces vortical in-plane displacements, and a normal-strain texture, as shown in Fig.~\ref{Fig2}(a) for a $3.15^\circ$ twist angle from first-principle calculations.
The normal-strain modulation of intralayer band gap produces a shallow moire potential for intralayer exciton (intra-X). Interlayer exciton (inter-X) is subject to a much deeper moire potential from coupling to the layer-contrasted piezoelectric charge background peaked at MX and XM with opposite signs (Fig.~\ref{Fig2}(b)). 
From the lowest intra-X and inter-X minibands, we construct {\it maximally localized Wannier functions} centered at MX and XM sites respectively (c.f. Fig.~\ref{Fig2}(d) and Supplementary Note~2). 
The deep moire potential for inter-X confines its center-of-mass (COM) motion into compact S-type and P-type orbitals (referred to as inter-S and inter-P), with a large energy separation. 
In contrast, intra-X features rather extended S- and P- orbitals. 

Fig.~\ref{Fig2}(c-d) shows the energy level schemes of the moire exciton orbitals at this twist angle. At zero displacement field ($E_z =0$), the intra-S and intra-P levels lie between the widely separated inter-P and inter-S levels. Their hybridization is field-tuned through the Stark effect of inter-X, and the P-wave form of the interlayer tunneling matrix element $T(\bf r)$ near MX/XM dictates that inter-S (inter-P) hybridizes only with intra-P (intra-S)~\cite{zheng_forster_2025,Hongyi2021PRX}.
The hybridization branch consisting of intra-S and inter-P is the one accessible via normal-geometry absorption and photoluminescence, through the bright intra-S component.
The ground state, primarily consisting of inter-S, is essentially invisible under these circumstances because it can only hybridize with the nearly dark intra-P. 
The phase texture of $T(\bf r)$ yields an intra-P component with azimuthal quantum number $m=-\tau$ ($\tau=\pm1$ being the valley index, details in Supplementary Note~7).

In an array of such hybrid moir\'e excitons, cooperative radiation and non-radiative F\"orster-type coupling can be treated on equal footing using the dipole-dipole coupling operator $\hat{G}_{\mathrm{dd}}
\equiv
\frac{\omega^{2}}{\epsilon_{0}c^2}\hat{\mathbf P}^{\dagger}(\mathbf r_{1})
\cdot
\overleftrightarrow{G}(\omega,\mathbf r_{1}-\mathbf r_{2})
\cdot
\hat{\mathbf P}(\mathbf r_{2})$~\cite{Asenjo_PRX2017_pGp,HuangPRL2025_moire_radiation}, where $\overleftrightarrow{G}$ is the Dyadic Green's tensor, and $\hat{\mathbf P}$ denotes the dipole operator. 
Its matrix element between the Wannier basis states can be separated into Hermitian and anti-Hermitian kernels,
\begin{equation}
\langle\Psi^{\rm intra}_{m',\tau'}(\mathbf R_j)|\hat G_{\rm dd}|\Psi^{\rm intra}_{m,\tau}(\mathbf R_i)\rangle\\
 =\mathcal J_{m,\tau}^{m',\tau'}(\mathbf r)
 +\frac{i}{2}\Gamma_{m,\tau}^{m',\tau'}(\mathbf r).
 \label{eq:wavepacketGreen}
\end{equation}
with $r=\mathbf R_i-\mathbf R_j$. 
In the evanescent near-field limit, the Hermitian kernels $\mathcal J_{m,\tau}^{m',\tau'}(\mathbf r)$ give the F\"orster dipole-dipole coupling~\cite{zheng_forster_2025}. 
The intravalley and intervalley channels satisfy $\mathcal J_{1,-1}^{-1,1}(\mathbf r)=-\mathcal J_{1,1}^{1,1}(\mathbf r)$ (Supplementary Note~4). The hermitian kernels are therefore diagonalized in the valley coherent basis: $|\Psi^{\rm intra}_{p,{\rm TE}}\rangle=\frac{1}{\sqrt{2}}\left(|\Psi^{\rm intra}_{-1,K}\rangle+|\Psi^{\rm intra}_{1,K'}\rangle\right)$ and $|\Psi^{\rm intra}_{p,{\rm TM}}\rangle=\frac{1}{\sqrt{2}}\left(|\Psi^{\rm intra}_{-1,K}\rangle - |\Psi^{\rm intra}_{1,K'}\rangle\right)$. 
The on-site kernel shifts the TM state upward in energy by $\sim 70$ meV (Fig.~\ref{Fig2}(c)), while for the TE state, the Hermitian kernels vanish for both on-site and inter-site couplings.
The intra-P TE state is therefore the closest one in energy for hybridization with the inter-S state, leading to the lowest-energy moire exciton trapped at each individual site $i$,
\begin{equation}
 |X_{i}\rangle = \sqrt{1-\mathcal W}|\Psi^{\rm inter}_{s}(\mathbf R_{i})\rangle
 +\sqrt{\mathcal W}|\Psi^{\rm intra}_{p,{\rm TE}}(\mathbf R_{i})\rangle.
 \label{eq:hybridX}
\end{equation}
Hole tunneling dominates the interlayer hybridization~\cite{zheng_forster_2025,WangYong2017PRB}. The resultant layer configurations of the hybrid exciton at the two sublattices are shown in insets of Fig.~\ref{Fig3}(b). 
The only free parameter for determining $\mathcal W$ is the binding energy difference between intra-X and inter-X, $\Delta E_b$, taken to be $72$~meV in Fig.~\ref{Fig2}(d), yielding $\mathcal W=0.12$ at zero displacement field. 
Microscopic calculations for TMD bilayers find values ranging from several tens to $\sim$100 meV, exhibiting a pronounced dependence on the surrounding dielectric~\cite{Ovesen2019InterlayerDynamics,Torun2018InterIntraExcitons,Kamban2020InterlayerBinding}, which can be exploited to tune $\mathcal W$. 
The anti-Hermitian kernel of $\hat{G}_{dd}$ encodes the radiative property, whose projection onto the above hybrid exciton wavepacket is,
\begin{equation}
 \Gamma_X(\mathbf r)=\mathcal W\Gamma_p(\mathbf r)
 =\mathcal W \left[
 \Gamma_{1,1}^{1,1}(\mathbf r)
 +\Gamma_{1,-1}^{-1,1}(\mathbf r)
 \right],
 \label{eq:GammaXProjection}
\end{equation}
The onsite value $\Gamma_X \equiv \Gamma_X(\mathbf r=0)   =\frac{4}{3} \beta^4  \mathcal W \Gamma_0$ gives the radiative decay rate of the hybrid exciton at an individual trapping site. Here, $\Gamma_0 \sim O(1)$ meV is the decay rate of an intralayer exciton momentum eigenstate at the light cone center~\cite{moody_intrinsic_2015}, and $\beta \equiv k_0w \ll 1$, $w$ the width of the intra-P wavepacket. The $\beta^4$ dependence reflects the small weight of the intra-P wavepacket within the light cone of size $k_0 \equiv \frac{\omega}{c}$. The emitted photon is in an azimuthally polarized vector-vortex mode carrying one unit of OAM (see Fig.~\ref{Fig3}(a)),
with angular intensity distribution $\propto \sin^2{\theta}$, where $\theta$ is the polar angle.
The single-site emission rate, $\Gamma_X/\hbar \sim 10$ ms$^{-1}$, is practically negligible.
Nonetheless, unlike the short-ranged Hermitian kernel, $\Gamma_X(\mathbf r)$ has a spatial range set by the optical wavelength~\cite{Xu2026CooperativeOptics}, which contains $\sim (2\pi)^2(k_0a_M)^{-2} \sim O(10^4)$ moire exciton emitters -- precisely the regime where cooperative radiation effects become significant. 

\textit{Superradiant Bloch states of excitons and biexcitons}. The extended intra-P orbital component of the hybrid exciton facilitates kinetic hopping between next-nearest-neighbor sites (i.e., within each sublattice), whereas hopping between sublattices is forbidden by their opposite layer configurations (c.f. Fig.~\ref{Fig3}(b)). The single-exciton eigenstates are therefore Bloch states on each sublattice. Their radiation properties are obtained by projecting the anti-Hermitian kernel $\Gamma_X(\mathbf r)$ onto the Bloch basis (Supplementary Note~5). This projection encodes the cooperative radiation between different moir\'e sites (Fig.~\ref{Fig3}(b)), and the phase-matched radiation amplitudes add coherently, enforcing conservation of in-plane momentum $\mathbf{k}$, which determines the emission polar and azimuthal angles $\theta$ and $\phi$: $\mathbf{k}=k_0(\sin{\theta}\cos{\phi},\sin{\theta}\sin{\phi})$.
In sharp contrast to the superradiant Bloch states of structureless emitters~\cite{Xu2026CooperativeOptics}, the P-orbital nature of the intra-X component renders the Bloch state at $\mathbf{k}=0$ completely dark. The radiative decay rate strongly depends on $k$, or equivalently the polar angle $\theta$: $\widetilde\Gamma_X (\theta)=\mathcal W\Gamma_{0}\frac{\sin^2\theta}{\cos\theta}\frac{1}{\sum_{\mathbf G}(G/k_0)^2 e^{-w^2G^2}}$, $\mathbf G$ being the moir\'e reciprocal lattice vectors, as plotted in Fig.~\ref{Fig3}(c). See details in Supplementary Note~5. At $k \sim \frac{1}{2}k_0$, the Bloch state is already strongly superradiant with $\widetilde{\Gamma}_X \sim 10^4 \Gamma_X$, corresponding to a radiative lifetime of $\sim O(10)$ ns. 

We next turn to the biexciton. The strong inter-sublattice dipolar attraction binds a pair of excitons on a nearest-neighbor bond, giving rise to three quadrupolar biexciton (XX) configurations depending on the bond orientation. The excitonic kinetic hopping (Fig.~\ref{Fig3}(b)) couples these configurations to form Bloch states on an effective kagome geometry~\cite{NSR_Wang_2026}. 
The biexciton emission is dominated by the on-shell time-ordered cascaded process involving an intermediate recoil-exciton, which is accurately captured by the anti-Hermitian part of $\hat{G}_{dd}$ (Supplementary Note~6).
The biexciton correlation significantly alters the cooperative radiation.
In contrast to the single-exciton Bloch state, the biexciton Bloch state is superradiant at $\mathbf{k}=0$, whose cascaded emission yields a momentum-entangled photon pair with the spatial-temporal wavefunction (c.f. Supplementary Note~6),
\begin{align}
\Psi(\theta;t_1,t_2)
 &=\Theta(t_2-t_1) \sin\theta \sqrt{\widetilde\Gamma_X(\theta) \, \widetilde\Gamma_{XX} }\nonumber\\
 &\times e^{-\widetilde\Gamma_{XX}(0)t_1/(2\hbar)}
 e^{-\widetilde\Gamma_X(\theta)(t_2-t_1)/(2\hbar)}\nonumber\\
 &\times e^{-i(\omega_{XX}+\omega_X)t_1}
 e^{-i\omega_X(t_2-t_1)}/\hbar. \label{eq:cascadeBiphoton}
\end{align}
Here $\Theta(t_2-t_1)$ is the heaviside step function, and $\hbar\omega_{XX}=\hbar\omega_X - \Delta$, where $\hbar\omega_X$ is the exciton energy and $\Delta$ is the biexciton binding energy.
The probability of emitting the photon pair at polar angle $\theta$ is also given by $\sin^2\theta$, where the linewidth of the two photons corresponds to that of the recoil exciton state $\widetilde\Gamma_X(\theta)$~\cite{AvanakiSchatz2021}.

The radiative decay rate $\widetilde\Gamma_{XX}$ of the biexciton Bloch state remains nearly constant over a momentum-space range well beyond the light cone (Fig.~\ref{Fig3}(e)).
The modest superradiant enhancement for a single state is due to the fact that the radiative kernel of $\hat{G}_{dd}$ can only couple biexciton configurations on bonds sharing a common site (Fig.~\ref{Fig3}(b)). As we show in the next section, such state acquires a macroscopic occupation in the condensate, further enhancing the entangled-photon-pair superradiance by the occupation number.

{\it Cooperative emission from coupled exciton and biexciton condensates.} 
We turn to the many-body ground state of hybrid moire excitons described by the extended Bose-Hubbard model, $ H=-t\sum_{\langle\!\langle ij\rangle\!\rangle}
 \left(b_i^\dagger b_j+{\rm h.c.}\right)
 +\frac{V_0}{2}\sum_i n_i(n_i-1)+\sum_{s=1}^{3}V_s\sum_{(ij)\in\mathcal B_s}n_in_j.$
Kinetic hopping occurs only between next-nearest-neighbors $\langle\!\langle ij\rangle\!\rangle$, i.e. within each sublattice only, as explained earlier. For the electrical dipole-dipole interaction,
we consider $V_0$ and $V_2$, the onsite and next-nearest-neighbor repulsion, respectively, and $V_1$ and $V_3$, the nearest-neighbor and third next-nearest-neighbor attraction, respectively.
The values of $t$ and $V_s$ depend on the moire period ($a_M$) and the degree of hybridization ($\mathcal W$), both controlled by the twisting angle. For the example of $3.15^\circ$ twisting, with $\mathcal W=0.12$ (Fig.~\ref{Fig2}(d)), we find $t=0.63, V_0=62, V_1=-6, V_2=1.8, V_3 =-1.4$ meV based on the Wannier orbitals. These energy scales are many orders of magnitude larger than the anti-hermitian radiative kernels $\Gamma_X$ and $\Gamma_{XX}$. Radiation can therefore be treated perturbatively, by evaluating the radiative kernels in the many-body ground state. The large onsite $V_0$ further justifies the hard-core limit to simplify calculations. 


The many-body ground state is obtained using density-matrix renormalization group calculations with cylinder boundary conditions, fixing $N_x=30$ and taking $N_y=4,6,8$.
From the correlation functions, $n_{X}(\mathbf k)= \sum_{\mu}\langle b_{\mathbf k,\mu}^\dagger b_{\mathbf k,\mu}\rangle$ and  
$n_{XX}(\mathbf k)=\langle B_{\mathbf k}^\dagger B_{\mathbf k}\rangle$ are extracted,
which are the exciton and biexciton populations respectively on momentum $\mathbf k$. $b_{\mathbf k,\mu}=N_\mu^{-1/2}\sum_{i\in\mu}e^{-i\mathbf k\cdot\mathbf R_i}b_i$  and $B_{\mathbf k}=N_b^{-1/2}\sum_{\langle ij\rangle}e^{-i\mathbf k\cdot\mathbf (\mathbf R_{i}+\mathbf R_{j})/2}b_{i} b_{j}$, where $N_{\mu}$ is the number of sites on sublattice $\mu$ and $N_b$ is the number of nearest-neighbor (MX-XM) bonds. Figure~\ref{Fig4}(a) shows that 
both $n_{X}(\mathbf k)$ and $n_{XX}(\mathbf k)$ develop a macroscopic peak centered at $\mathbf k=0$. The width of the peak is due to the finite size calculations.
Fig.~\ref{Fig4}(b) shows the condensates populations as functions of intra-X weight $\mathcal W$, tunable through the displacement field. Figure~\ref{Fig4}(c) further indicates that both $n_X(\mathbf k =0)$ and $n_{XX}(\mathbf k =0)$ are extensive quantities, increasing linearly with $N_y$ and diverging in the thermodynamic limit $N_y\rightarrow\infty$.
This macroscopic occupation of zero-momentum Bloch states is the hallmark of coupled exciton and biexciton condensates (Supplementary Note~8). 

With the macroscopic occupation of the superradiant biexciton Bloch state, the condensate decays through superradiating the momentum-entangled photon pair (Eq.~(\ref{eq:cascadeBiphoton})) with an emission rate of $n_{XX}(0) \widetilde \Gamma_{XX}/\hbar$.
In contrast, the condensate component $n_{\rm X}$ remains strictly dark at $\mathbf k =0$ (Fig.~\ref{Fig3}(c)). 
An in-plane gradient of the displacement field, $\nabla_{\parallel} E_z$, imprints a phase gradient on the atomic-type exciton condensate that is opposite on the MX and XM sublattices, owing to their opposite electric dipoles. This acts like a London-type acceleration term, driving a counterflow of exciton supercurrent: $\dot{\mathbf k}_{\rm MX/XM}= \pm\frac{d_0}{\hbar}\nabla_{\parallel} E_z$.
Fig.~\ref{Fig4}(a,d) shows the time evolution by applying the evolution operator $U$ corresponding to the Stark shift of on-site energy in $\nabla_{\parallel} E_z$ to the DMRG ground state. The $n_{\rm X}$ peak splits into two and displaced to finite momenta, whereas the $n_{\rm XX}$ peak remains unchanged as the quadrupolar biexciton is not coupled to $\nabla_{\parallel} E_z$.
The field gradient thereby activates the single-photon superradiant channel (Fig.~\ref{Fig3}(c)), which produces highly directional emission encoding the phase gradient in the atomic-type condensate component.
A small phase gradient of $\sim 0.02k_0$ is sufficient to switch the dominant condensate radiation channel from biexciton to exciton. At phase gradients $> k_0$, the exciton superradiance is quenched and biexciton radiation of entangled photon pairs again becomes the sole channel (c.f. Fig.~\ref{Fig3}(c,e)). 


{\it Outlook.} - The combination of electrically controlled, structured superradiance from atomic- and molecular-type condensates establishes t-MoTe$_2$ moire excitons as an intriguing platform in which cooperative quantum optics directly interrogates atomic-molecular many-body order.
From finite-temperature cluster mean-field calculations~\cite{You2012KagomeSuperfluidity}, a favorable BEC transition temperature of $\sim 2-14$ K is estimated for both exciton and biexciton condensates at $\mathcal W \sim 0.12 - 0.13$ (Supplementary Note~9), a temperature already achieved in existing t-MoTe$_2$ experiments~\cite{Cai2023FQAH,Zeng2023FCI,Park2023FQAH,Xu2023FQAH,Xu2025Superconductivity}.
Implementing oblique emission detection and displacement-field-gradient control would provide direct access to characteristic optical fingerprints of the atomic–molecular many-body order.


\begin{figure*}[p]\centering
\includegraphics[width=0.96\textwidth]{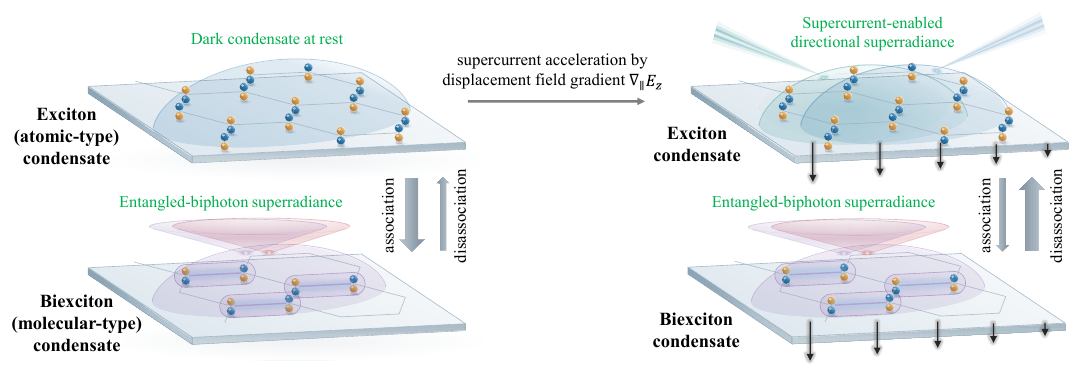}
\caption{{\bf Schematic of superradiance from co-existing exciton and biexciton condensates in twisted MoTe$_2$.} The biexciton condensate emits via entangled-biphoton superradiance. The exciton condensate is dark at zero momentum, whereas a supercurrent enables its directional superradiance. }
 \label{Fig1}
\end{figure*}

\begin{figure*}[p]\centering
\includegraphics[width=0.96\textwidth]{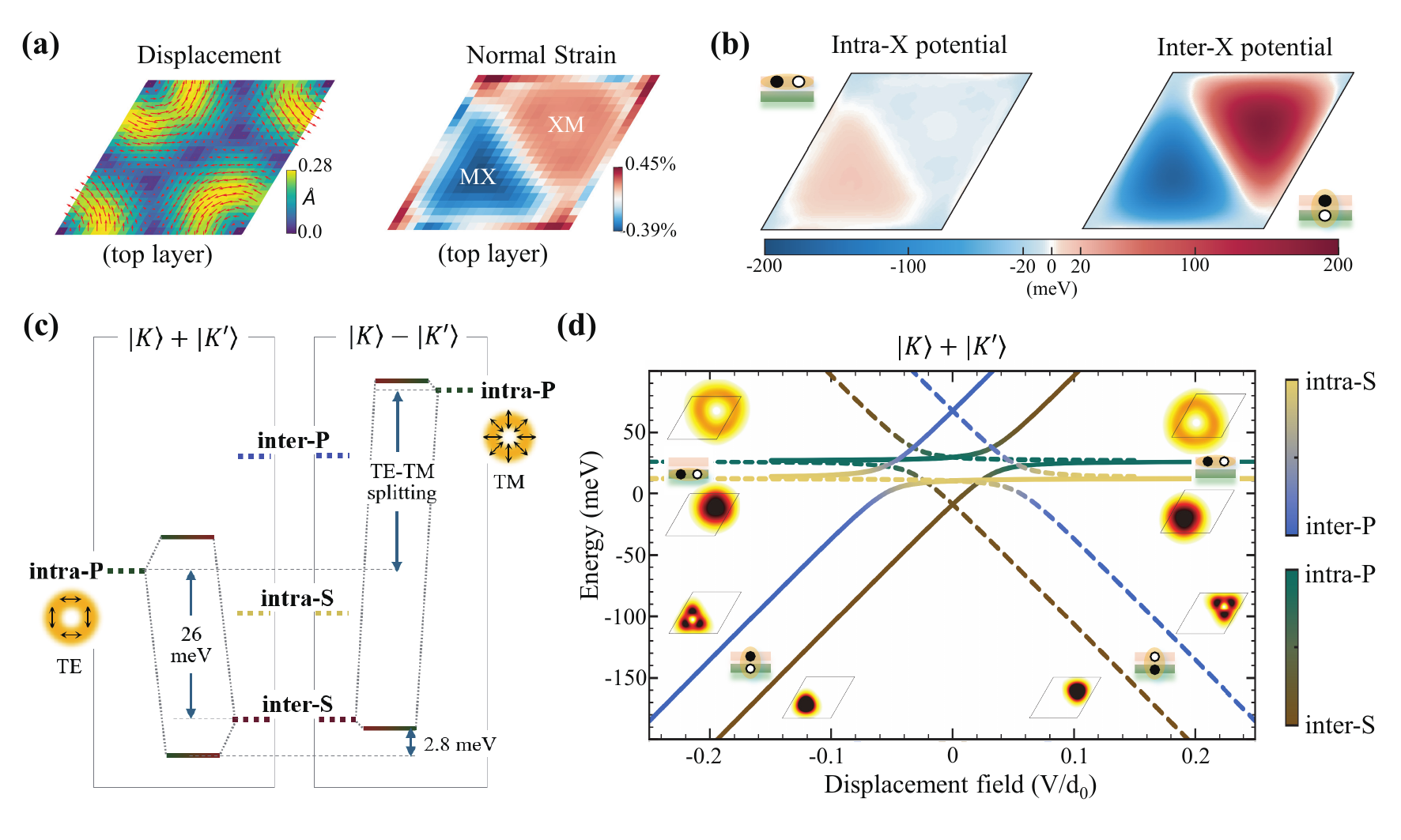}
\caption{\textbf{Hybrid moir\'e excitons in structural relaxed twisted MoTe$_2$.} \textbf{a}, In-plane displacement of Mo atoms and normal strain in the top layer of $3.15^\circ$ twisted bilayer MoTe$_2$. \textbf{b}, Moir\'e potentials for intralayer- and interlayer-excitons, generated by the strain-modulated band gap and polarization charge respectively (Supplementary Note~2). \textbf{c}, Zero-field level schemes for the TE ($|K\rangle+|K'\rangle$) and TM ($|K\rangle-|K'\rangle$) valley-coherent sectors, showing the symmetry-dictated Inter-S/Intra-P hybridization. \textbf{d}, Hybrid-exciton spectrum versus displacement  field in the TE valley-coherent sector. Solid and dashed curves correspond to opposite layer configurations, and colours encode the Inter-S, Inter-P, Intra-S and Intra-P character. Insets show their representative COM Wannier orbitals and layer configurations. }
 \label{Fig2}
\end{figure*}

\begin{figure*}[p]\centering
\includegraphics[width=0.9\textwidth]{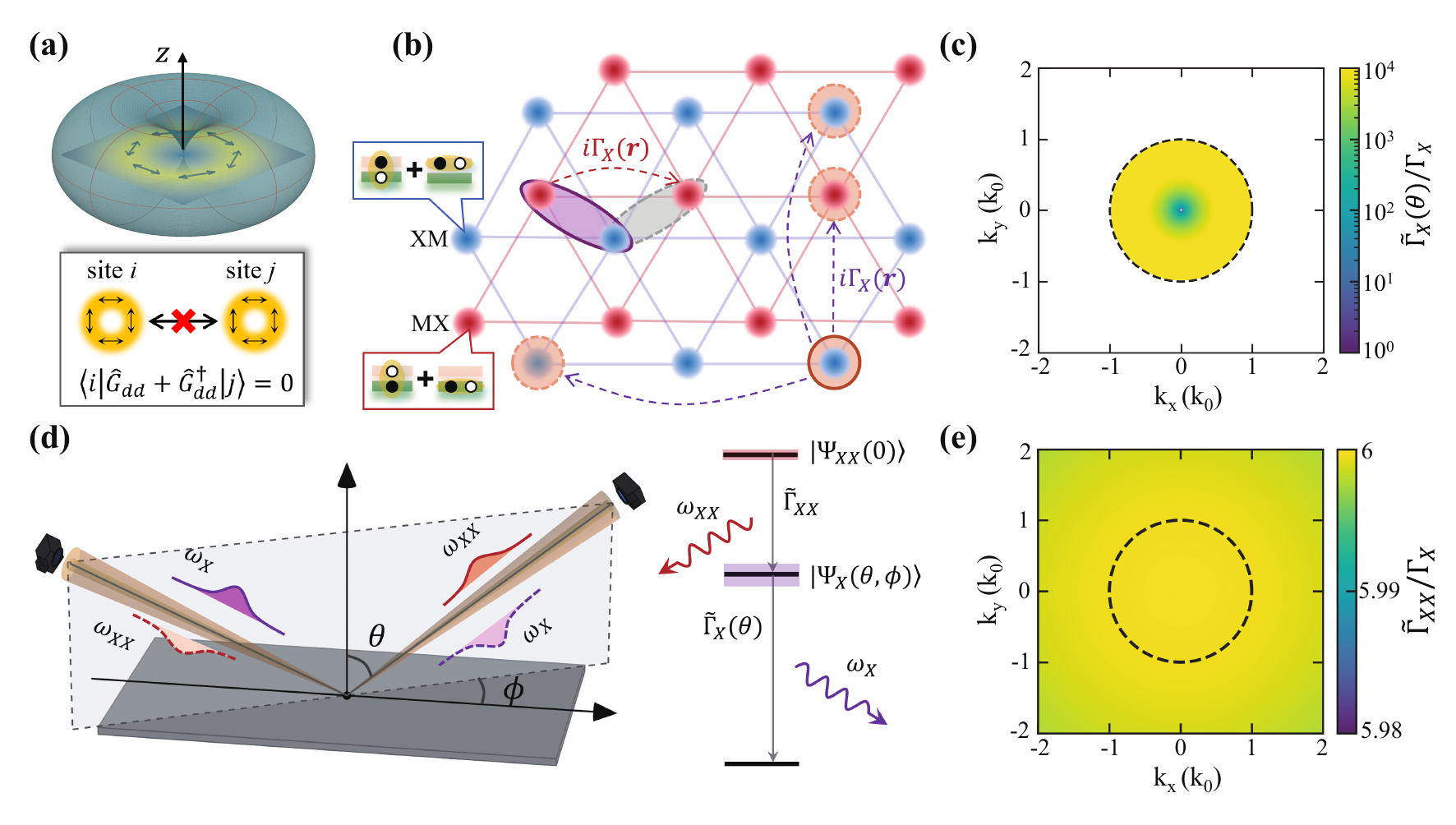}
\caption{\textbf{Cooperative radiation of excitons and biexcitons.} \textbf{a}, Structured light emission from the exciton ground state in an individual moir\'e site (c.f. Fig.~\ref{Fig2}c). The contour plot shows the angular distribution of emission intensity, with polarization indicated by arrows. Inset: for such TE-polarized OAM emitters, the non-radiative (Hermitian) part of resonant dipole-dipole coupling vanishes. \textbf{b}, Moir\'e exciton superlattice. Solid links denote the next-nearest-neighbour kinetic hopping $t$; inter-sublattice hopping is absent due to the opposite layer configurations(c.f. insets). Dashed arrows denote the radiative (anti-Hermitian) part of the dipole–dipole coupling, which links exciton sites over long distances, and can also link site-sharing biexciton configurations. \textbf{c}, Emission rate $\tilde{\Gamma}_X$ of exciton Bloch state, plotted as function of wavevector $\bf{k}$, in unit of the single-site emission rate $\Gamma_X$. $\tilde{\Gamma}_X$ depends only on $\theta = \arcsin{|{\bf k}| c/\omega }$, the polar angle of the emission. The Bloch states are subradiant near the light cone center ($\bf{k}=0$ state is completely dark), and superradiant elsewhere in the light cone. \textbf{d}, Cascade emission of path-entangled photon pair by a biexciton Bloch state. The first photon has a linewidth $\widetilde{\Gamma}_{XX}+\widetilde{\Gamma}_X(\theta)$ centered at $\hbar\omega_{XX} = \hbar\omega_X - \Delta$, and the second photon has linewidth 
$\widetilde{\Gamma}_X(\theta)$ centered at $\hbar\omega_X$. $\Delta$ is biexciton binding energy. \textbf{e}, Emission rate $\widetilde\Gamma_{XX}$ of biexciton Bloch state, plotted as function of wavevector $\bf{k}$, in unit of $\Gamma_{X}$. The Bloch state exhibits moderately superradiant emission of entangled photon pairs throughout the light cone and even beyond its boundary.}
 \label{Fig3}
\end{figure*}

\begin{figure*}[p]\centering
\includegraphics[width=0.9\textwidth]{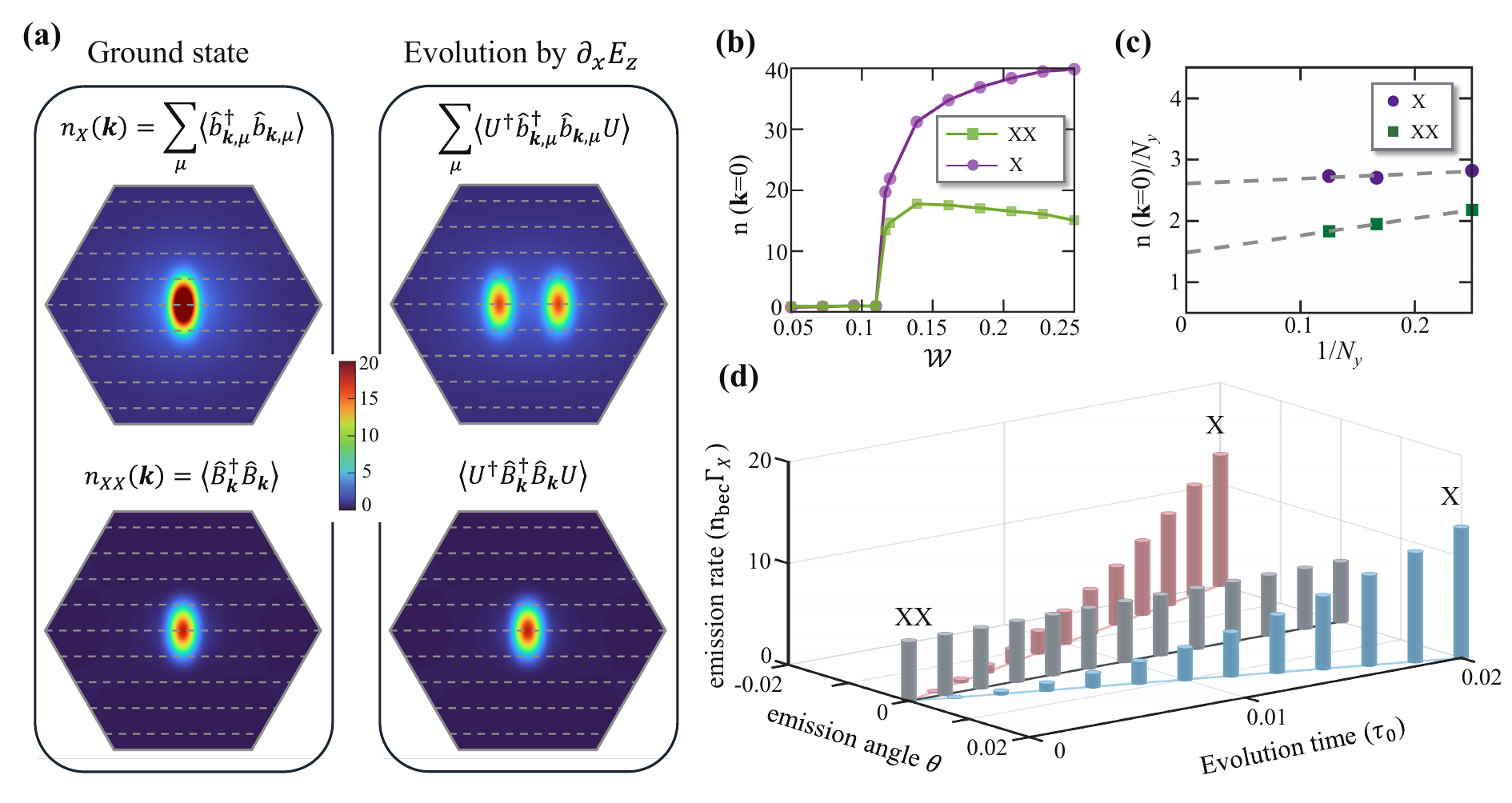}
\caption{\textbf{Coexisting exciton-biexciton condensates and electrically controlled superradiance.} \textbf{a}, Momentum-space occupations of exciton ($n_{X}({\bf k})$) and biexciton ($n_{XX}({\bf k})$) in the DMRG ground state (left column), and after evolution under an in-plane gradient of the displacement field (right column). Intralayer weight $\mathcal W=0.12$. \textbf{b}, Exciton and biexciton occupations at $\bf k =0$, as a function of $\mathcal W$. The DMRG calculations are performed on a cylinder of length $N_x=30$ and circumference $N_y=8$, at exciton filling factor of $\nu = 1/3$. \textbf{c}, Finite-cylinder scaling of the normalized $\bf k =0$ occupations. \textbf{d}, Emission rate and emission angle of the exciton condensates on the MX and XM sublattices, as functions of evolution time in the displacement field gradient. Time is given in units of $\tau_0=\hbar k_0/(d_0|\nabla E_z|)$. The emission rate of entangled biphoton from the biexciton condensate, which does not respond to the field gradient, is shown for comparison. The emission rates shown are angularly integrated and normalized by the sample area, expressed in units of $n_{\rm bec}\Gamma_{X}$, where $n_{\rm bec}$ is the density of condensed exciton/biexciton. In contrast to the biexciton superradiance, the supercurrent-enabled exciton superradiance is highly directional, with emission confined to a solid angle of $(Ak_0^2\cos{\theta})^{-1}$, $A$ being the sample area.}
 \label{Fig4}
\end{figure*}

{\it Acknowledgments.} - 
The work is mainly supported by the National Natural Science Foundation of China (No.12425406), Research Grant Council of Hong Kong (AoE/P-701/20, HKU SRFS21227S05), and New Cornerstone Science Foundation. H.W. also acknowledges support by NSFC Young Student Basic Research Program (No.125B2068). S.L. and M.C. acknowledge support by the National Natural Science Foundation of China (Grants No. 12174098 and No. 12574262), the Major Fundamental Research Program of Hunan Province (Grant No. 2025ZYJ004). Calculations were carried out in part using computing resources provided by 
Beijing PARATERA Tech Co., Ltd. (https://cloud.paratera.com) and by the High-Performance Computing Platform of Hunan Normal University. ITensor library~\cite{Fishman2022ITensor} is used in the DMRG simulations
\bibliography{Reference}

\end{document}